# K$_x$Fe$_{2-y}$Se$_2$ single crystals: Floating-zone growth, Transport and Structural properties

Y. Liu, Z. C. Li, W. P. Liu, G. Friemel, D. S. Inosov, R. E. Dinnebier, Z. J. Li, and C. T. Lin[*]

*Max Planck Institute for Solid State Research,*
*Heisenbergstraße 1, D-70569 Stuttgart, Germany*

Single crystals of superconducting K$_x$Fe$_{2-y}$Se$_2$ have been grown with the optical floating-zone technique under application of 8 bar of argon pressure. We found that large and high quality single crystals with dimensions of ~⌀6 × 10 mm could be obtained at the termination of the grown ingot through quenching, while the remaining part of the ingot decomposed. As-grown single crystals commonly represent an intergrowth of two sets of the *c*-axis characterized by slightly different lattice constants. Single crystal of K$_{0.80}$Fe$_{1.81}$Se$_2$ shows a superconducting transition at $T_c$ = 31.6 K, leading to a near 100% expulsion of the external magnetic field in magnetization measurements. On the other hand, neutron-diffraction data indicate that superconductivity in the sample coexists with a $\sqrt{5} \times \sqrt{5}$ iron-vacancy superstructure and static antiferromagnetic order. The anisotropic ratio of the upper critical field $H_{c2}$ for both *H*//*c* and *H*//*ab* configurations is ~3.46.

[*]Email address: ct.lin@fkf.mpg.de

## 1. Introduction

The discovery of superconductivity at temperatures above 30 K in $K_xFe_{2-y}Se_2$ compounds has attracted considerable attention [1]. The crystal structure of $K_xFe_{2-y}Se_2$ descends from the well-known $ThCr_2Si_2$ structure with vacancies at the Fe sites. It was soon found that magnetism and superconductivity coexist in $A_xFe_{2-y}Se_2$ ($A$ = K, Rb, and Cs) [2-3, 13]. Even in single crystals which show full shielding of an external magnetic field below the superconducting transition temperature ($T_c$), antiferromagnetic order is found below the Néel temperature ($T_N$), as high as 500 to 540 K, depending on the alkaline element $A$ [3]. The antiferromagnetic transition is coupled with a structural transition of iron vacancy ordering. The relation between superconductivity and iron-vacancy ordering structure remains a subject of debate. It has been found that the $K_xFe_{2-y}Se_2$ system exhibits a transition from an insulating state to a superconducting state upon variation of the iron content [4]. In fact, the stoichiometric $K_{0.8}Fe_{1.6}Se_2$ compound with ideal iron vacancy ordering structure shows insulating behavior. Moreover, microstructure analysis by means of transmission electron microscopy (TEM) on $K_xFe_{2-y}Se_2$ single crystals demonstrates a clear phase separation in the superconducting samples. Two phases in the form of parallel lamellae, namely iron-vacancy-ordered and -disordered phases, stack along the $c$ axis of the crystal [5]. Furthermore, the temperature dependence of single-crystal x-ray diffraction (XRD) in transmission mode, using synchrotron radiation, reveals a minority non-magnetic phase with a weak $\sqrt{2}\times\sqrt{2}$ superstructure, in addition to the majority magnetic phase with a $ThCr_2Si_2$-type tetragonal lattice modulated by the $\sqrt{5}\times\sqrt{5}$ iron-vacancy ordering below 520 K [6].

These complex microstructures make it difficult to grow bulk superconducting $A_xFe_{2-y}Se_2$ ($A$ = K, Rb, and Cs) single crystals. Success in obtaining superconducting samples strongly depends on the ratios of the starting chemical components, particularly on the initial Fe content, as well as heating, cooling and growth rates. So far, different techniques such as flux growth and the Bridgman method have been used to grow $A_xFe_{2-y}Se_2$ (A=K, Rb, and Cs) single crystals [3-4,7-12] with typical procedure as follows: First, FeSe precursor was synthesized at 700 °C in an evacuated quartz tube. Then, FeSe precursor was ground into powder and mixed with K(Rb,Cs) at a ratio of K(Rb,Cs):Fe:Se=0.8:2:2. The mixtures were loaded into a double sealed quartz ampoule and then heated to 1030 °C and kept at this temperature for 2 h. A fast cooling rate of 6 °C/h was applied before turning off the furnace at 750 °C. However, the as-grown crystals obtained by these methods hardly showed 100% superconducting volume, mainly due to the inhomogeneity of the samples, since ordered or disordered vacancy phases could be randomly formed due to an uncontrolled thermal distribution of iron vacancies during growth under the described conditions. In this study, we report that superconducting $K_xFe_{2-y}Se_2$ single crystals can be reproducibly grown by the optical floating-zone (OFZ) technique. Single crystals of a large size and high quality and high superconducting volume fraction are achievable, offering sufficient material for future inelastic neutron-scattering experiments.

## 2. Experimental details

The raw materials for the preparation of a feed rod were K (99.95%), Fe (99.995%) and Se (99.999%). The feed rod was synthesized by the one-step solid-state reaction method. Elemental K, Fe and Se were weighed at a ratio of K:Fe:Se=0.8:2:2, and loaded into an alumina crucible. It should be emphasized that K and Se were separated by Fe during loading, preventing K from strongly reacting with Se. The elements were then sealed in a quartz tube under 400 mbar argon atmosphere and heated to 850 °C for 10 hours. Finally, the sintered $K_{0.8}Fe_2Se_2$ mixture was ground into a powder. The powder was pressed into a feed rod with a cylindrical shape ~6–7 mm in diameter and ~70–80 mm in length with 600 bar of hydrostatic pressure. The feed rod was used to grow the crystal directly without the conventional process of sintering and pre-melting. A seed rod of 2 cm in length was cut from the feed rod.

The single crystal was obtained in an OFZ furnace with 4×300 W halogen lamps (Crystal System Inc. FZ-T-10000-H-III-VPR). The seed and feed shafts were rotated in opposite directions at rates of 20 rpm. The traveling rate was ~1.0 mm/h. Argon atmosphere was applied during the growth process at 8 bar. The composition of the crystals was determined by energy-dispersive X-ray spectroscopy (EDX). X-ray diffraction analysis was performed on a Philips Xpert XRD diffractometer using Cu Kα radiation λ=1.54056 Å. In-plane resistivity measurements were performed on a Physical Property Measurement System (PPMS, Quantum Design). DC magnetic susceptibility was measured using a SQUID-VSM magnetometer (Quantum Design). Neutron diffraction experiments were carried out to characterize the quality of as-grown single crystals at the IN3 instrument (ILL, Grenoble).

## 3. Results and discussion

The floating zone method enables us to obtain large and high quality single crystals with a mass of up to ~1.8 g. A typical as-grown single-crystal ingot of $K_xFe_{2-y}Se_2$ is shown in Fig. 1(a). As can be seen in Figs. 1(b) and (c), the cleaved termination of the ~1 cm ingot displays a large crystal grain with a shiny surface, while the remaining portion of the ingot appears to be non-crystalline with a bubble-like structure from the beginning towards the termination. This non-crystalline area has decomposed into K-poor K-Fe-

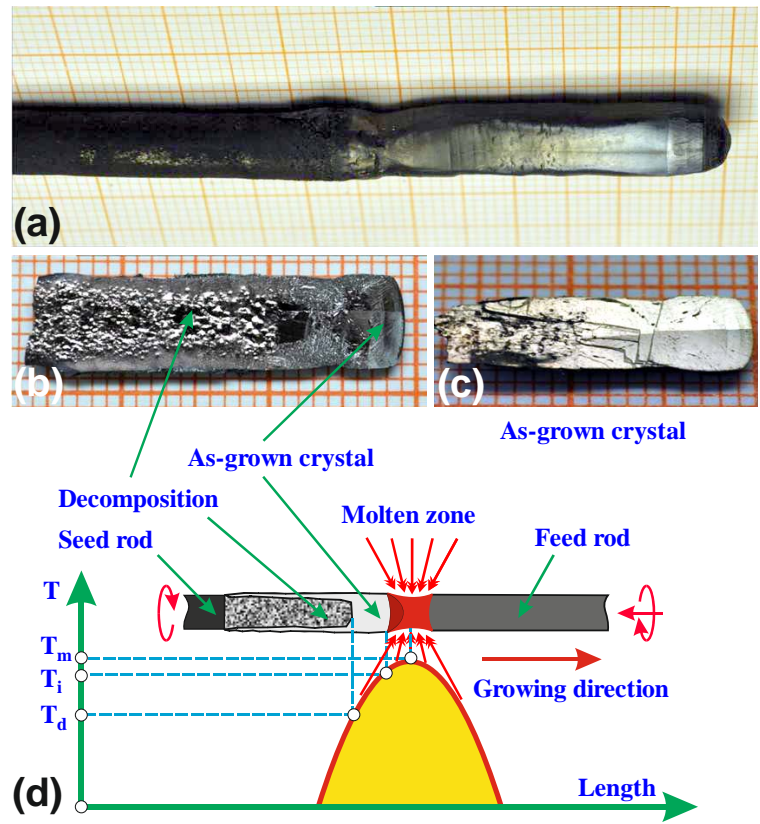

Fig. 1 (a) As-grown $K_xFe_{2-y}Se_2$ single crystal ingot. (b) and (c) single crystals were cut off from the ingot (a) and cleaved along the growing direction. A shiny surface with (00*l*) orientation can only be observed at the termination of the ingot, whereas the remaining part of the ingot had decomposed. (d) The schematic drawing illustrates the growth process of the $K_xFe_{2-y}Se_2$ single crystals, which form between the solid-liquid interface temperature, $T_i$, and decomposition temperature, $T_d$, below which the crystals undergo a continuous decomposition during crystal traveling.

Se, as determined by EDX. We also noticed thin filament-like deposits on the surface of the ingot throughout the entire growth process. This substance is pure Fe, according to EDX analysis. We have also observed that part of the grown ingot underwent decomposition during floating-zone growth. This feature strongly suggests that the $K_xFe_{2-y}Se_2$ single crystal is quite unstable. As the molten zone moves up, the already formed crystal continuously passes through a temperature zone, $T_d$, where the crystal begins to decompose, as illustrated in Fig. 1(d). Thus, to avoid passing through the decomposition zone the lamp power was switched off towards the end of the growth process, resulting in non-decomposed crystal phase at the terminal part of the ingot. Using a thermocouple for direct measurement of the temperature distribution below the molten zone, we observed accurately that $T_m$~889, $T_i$~784 and $T_d$~280 °C, for the molten zone, solid-liquid interface and decomposition temperatures, as illustrated in Fig. 1(d), respectively. Therefore, as-grown crystals can only be obtained by quenching at between ~784 and ~280 °C. We also observed that a stable molten zone can be formed by optimizing the growth parameters of rotation and traveling velocity. In this study, the superconducting $K_xFe_{2-y}Se_2$ single crystal could be grown at a traveling velocity of ~0.8 mm/h under 8 bars of argon pressure. The rotation of seed and feed rods is another important factor in producing a molten zone with a homogeneous composition distribution, which is a prerequisite condition for forming high quality single crystals.

It is interesting to note that the inner diameter of ~∅5 mm of the grown ingot decomposed at ~280 °C, while an ingot ring width of about 1–1.5 mm remained in single crystal form, as shown in Fig. 1(b) and illustrated by Fig. 1(d), respectively. This phenomenon indicates that the decomposition is caused by a great thermal strain occurring in the inner crystal ingot. The fact is that after the crystal formed the latent heat of crystallization was hardly released along the radial direction which corresponds to the robust (001) layers, perpendicular to the growth direction of the ingot. Therefore the compound decomposed to a K-poor K-Fe-Se.

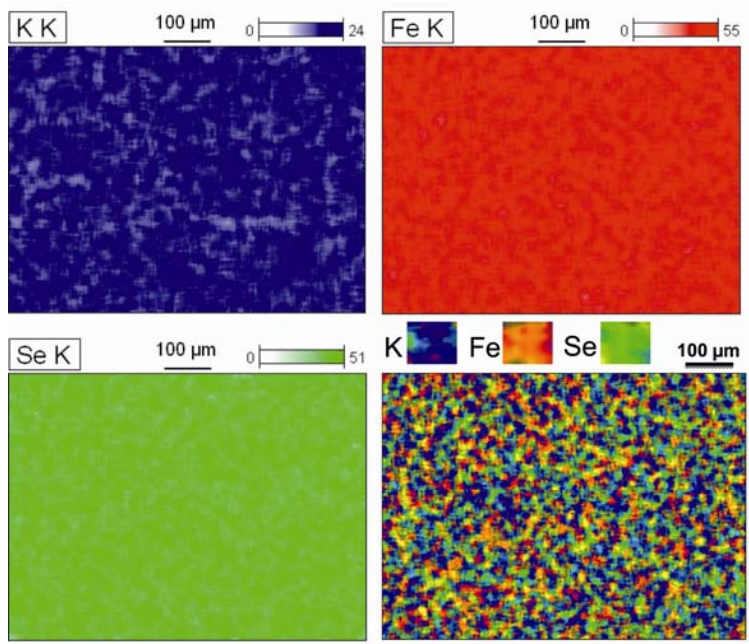

Fig. 2 Composition in the *ab* plane was determined by EDX with a mapping function as K:Fe:Se=0.80:1.81:2. Three elements are homogeneously distributed and overlapped in the same area shown in the four panels, respectively. The yellow spots are formed mainly by the overlapping of blue and green.

Black plate-like crystals with a shiny surface were cleaved from the termination of the crystal rod as shown in Fig. 1(c). By normalizing the Se content as a mol number with 2, Fe content slightly fluctuates between $1.81 < 2-y < 1.89$, while K between $0.74 < x < 0.8$ for the crystals cleaved from the different area of the end portion about 1 cm in length. A typical single crystal with composition of $K_{0.80}Fe_{1.81}Se_2$ is used in this study. Figure 2 reveals the homogeneous distribution of K, Fe and Se elements in these samples. Furthermore, it should be noted that the iron content in our many samples is approximately 1.81, which strongly deviates from the iron content of 1.6 that is most favorable for the formation of an iron-vacancy-ordered phase. A high variation of Fe content was also observed between $1.60 < 2-y < 1.79$ in the crystals grown under the Ar flowing condition.

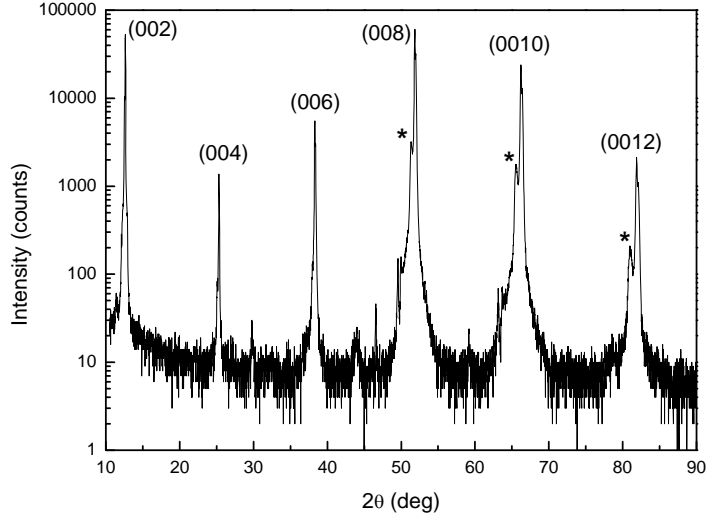

Fig. 3. XRD pattern of $K_{0.80}Fe_{1.81}Se_2$ single crystal. The shoulders located beside the main reflections are indicated by the asterisks, which correspond to the possible superconducting phase.

The single crystals commonly represented an intergrowth of the iron-vacancy-ordered states and -disordered states along the *c*-axis, which are characterized by slightly different lattice constants. Our XRD pattern was obtained by the (00*l*) reflections of the $K_{0.80}Fe_{1.81}Se_2$ single crystal, consistent with previous results [1,11], as shown in Fig. 3. Three shoulders besides the (008), (0010) and (0012) reflections can be discerned, as indicated by the asterisks. This second set of the (00*l*) reflections is related to the phase separation phenomenon [11]. Single crystal powder XRD was performed to interpret the structure space group belonging to I4/m, related to the superstructure with the stronger reflections. The stronger reflections correspond to the lattice parameters of $a = b = 8.688$ and $c = 14.128$ Å that are commensurate with the superstructure reflections, indicating that they originate from the vacancy-ordered insulating phase. Therefore, the second phase with $c = 14.134$ Å (asterisks) must be attributed to the metallic phase without iron-vacancy ordering. The recent study of μSR indicates that only a small volume (10%) of the sample is responsible for the superconducting phase [12]. Therefore, we assume that the longer *c*-axis of 14.134 Å should be responsible for the superconducting phase with the iron-vacancy-disordered states, while the shorter one with a *c*-axis of 14.128 Å being responsible for the superstructure phase with the iron-vacancy-ordered states.

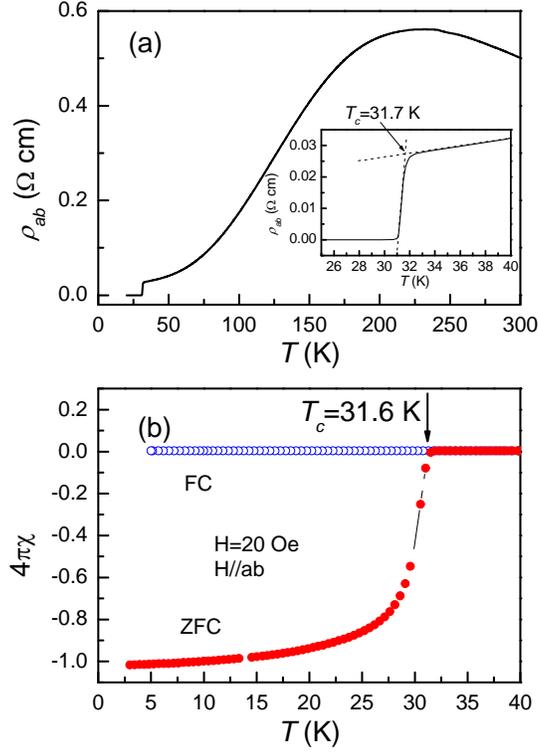

Fig. 4 (a) Temperature dependence of in-plane resistivity $\rho_{ab}$ of a $K_{0.80}Fe_{1.81}Se_2$ single crystal. Inset shows a superconducting transition temperature at around $T_c = 31.7$ K. (b) Magnetization curves show the superconducting transition temperature at $T_c = 31.6$ K.

The temperature dependence of the in-plane resistivity $\rho_{ab}$ of the $K_{0.80}Fe_{1.81}Se_2$ single crystal is shown in Fig. 4 (a). A clear hump is observed around 220 K. This hump can be shifted by varying Fe content in $K_xFe_{2-y}Se_2$ single crystals, which has been interpreted in terms of the ordering process of the cation vacancies in the nonstoichiometric $K_xFe_{2-y}Se_2$ [4]. When considering the phase separation scenario, however, we suggest that the resistivity hump could result from the competition between the two electronic transport channels, i.e., insulating (iron vacancy ordering) and superconducting phases (iron vacancy disordering). Below 220 K, $\rho_{ab}$ shows metallic behavior and the sample enters the superconducting state at $T_c = 31.7$ K. Consistent with the resistivity data, magnetization curves show the superconducting transition temperature at $T_c = 31.6$ K, leading to a near 100% expulsion of the external magnetic field in zero field cold below 5 K, as shown in Fig. 4 (b).

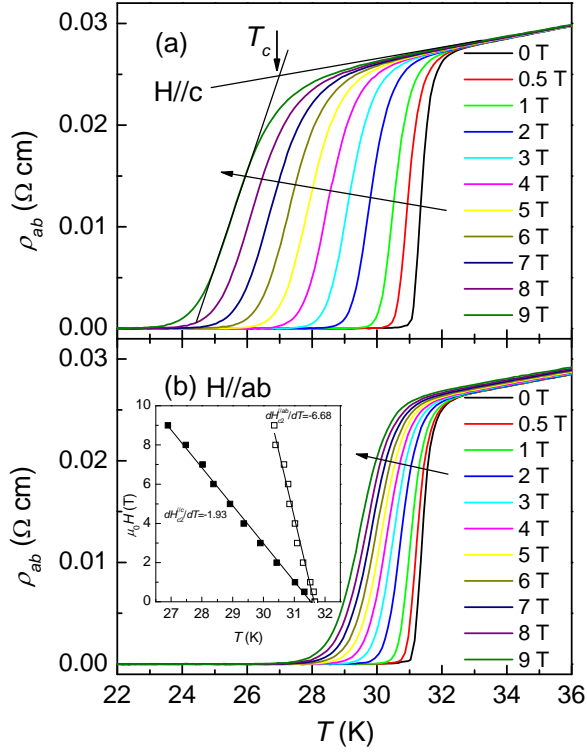

Fig. 5 The resistive broadening curves for the $K_{0.80}Fe_{1.81}Se_2$ single crystal under both (a) $H//c$ and (b) $H//ab$ configurations. The Inset of Fig. 5(b) shows the Vortex phase diagram of the single crystal. Solid lines represent fits with a linear relation to the upper critical field $H_{c2}$.

The resistivity broadening curves of the $K_{0.80}Fe_{1.81}Se_2$ single crystal with the magnetic field applied parallel to the $c$ axis ($H//c$) and with the field within the $ab$ plane ($H//ab$) of the sample, as shown in Figs. 5(a) and (b), respectively. For $H//c$ $\rho_{ab}$ dramatically broadens and fans out with increasing magnetic field, whereas it only shifts ~3 K towards the low temperature regime for $H//ab$. $T_c(H)$ was taken as the intersection of the linear extrapolation of normal state resistivity and the superconducting transition curve. The obtained upper critical fields $H_{c2}^c$ and $H_{c2}^{ab}$ are shown in the inset of Fig. 5(b). It can be seen that the upper critical fields $H_{c2}$ extracted from the resistivity measurements for the two field orientations display a nearly linear behavior with decreasing temperature in the vortex phase diagram for the single crystal. The slopes are deduced from a linear fitting to $H_{c2}(T)$ as $dH_{c2}^{ab}/dT = -6.68$ for $H//ab$ and $dH_{c2}^c/dT = -1.93$ for $H//c$, respectively. The anisotropic ratio of the slopes of $H_{c2}$ for the two field orientations reaches ~3.46, which is close to the 3.6 reported by Mizuguchi *et al.* [13].

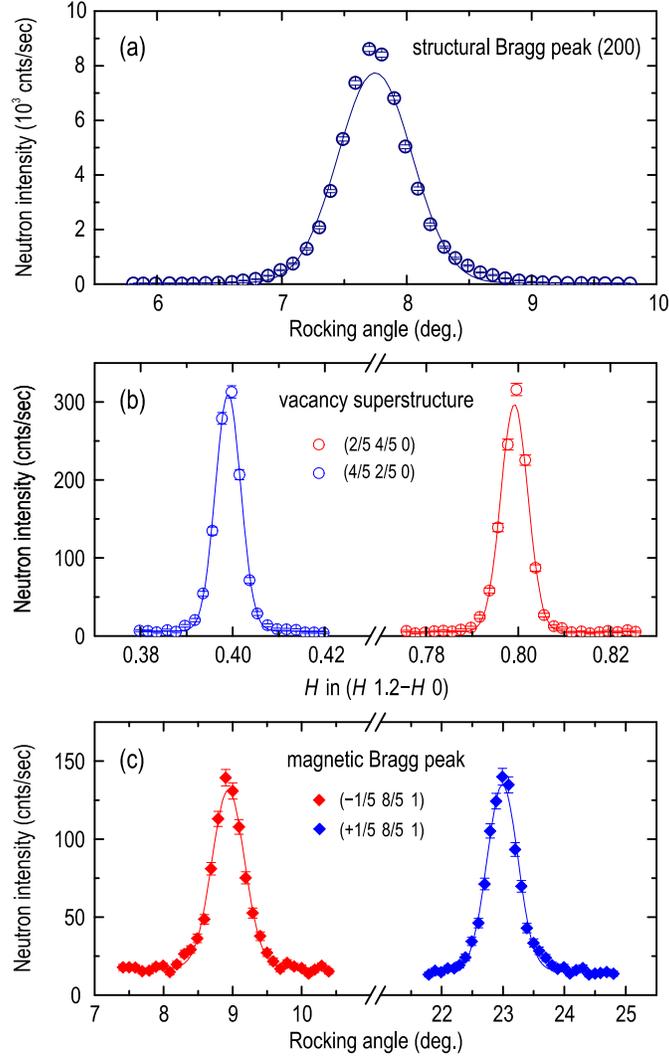

Fig. 6 Neutron diffraction data. (a) Rocking curve of the (200) structural Bragg peak, indicating a resolution-limited mosaicity of the single crystal. (b) Superstructure Bragg reflections originating from the $\sqrt{5} \times \sqrt{5}$ iron-vacancy ordering. The two peaks correspond to the twin domains with opposite orientation of the vacancy superstructure with respect to the parent lattice directions. (c) Magnetic Bragg reflections from the two twin domains, measured at room temperature. The solid lines in all panels are Gaussian fits.

Now we turn to the description of magnetism in our samples, which we have characterized using single-crystal neutron diffraction. The data presented in Fig. 6 have been acquired on one piece of the as-grown sample with a mass of ~100 mg. The neutron wavevector was set to $k_i$ = 2.662 Å$^{-1}$. The sample was mounted in the (*HK*0) scattering plane and could be additionally tilted using the instrument's goniometers in order to reach the out-of-plane magnetic Bragg reflections. The resolution-limited rocking curve measured on the (200) structural Bragg peak, shown in Fig. 6 (a), indicates that the mosaicity of the sample in the ab plane is of the order of ~0.5° or better.

As we already noted, the superconducting phase in the K$_x$Fe$_{2-y}$Se$_2$ compound typically coexists with insulating vacancy-ordered regions that exhibit strong antiferromagnetic ordering at room temperature. In Fig. 6 (b), we show two superstructure reflections originating from the $\sqrt{5}\times\sqrt{5}$ Fe-vacancy ordering. The two peaks at (2/5 4/5 0) and (4/5 2/5 0) correspond to different twin domains with the counterclockwise and clockwise rotation of the vacancy superstructure about the *c*-axis with respect to the parent lattice directions in the *ab* plane, respectively [14]. Their equal intensity indicates that the twins with both orientations are present in equal proportions in the sample. The presence of an antiferromagnetic order associated with this kind of vacancy superstructure, present at room temperature [2-3, 15], is evidenced by Fig. 6 (c). Here, we show the (±1/5 8/5 1) magnetic Bragg peaks from the two twin domains, measured at *T* = 300 K. Due to the strong magnetic moment that, according to some reports, reaches 3.3 μ$_B$/Fe [16-17], their intensities are very strong despite the nearly twofold suppression due to the iron form-factor at this large |**Q**| [14]. If the correction for the magnetic form-factor is taken into account, the resulting magnetic Bragg intensity turns out to be comparable to that of the structural reflections in panel Fig. 6 (b).

**4. Conclusions**

The optical floating zone technique can be successfully employed to grow the recently discovered K$_x$Fe$_{2-y}$Se$_2$ superconducting single crystals. To avoid decomposition of the as-grown single crystals a quenching treatment is applied at the temperature range between 784 and 280 °C. Both metallic and superstructure phases are observed by the XRD reflections. Resistivity and magnetic measurements show a superconducting transition at $T_c$ ~32 K. The anisotropic ratio of $H_{c2}$ for the two field orientations, *H*//*c* and *H*//*ab*, reaches ~3.46.

**Acknowledgements**

The authors would like to thank G. Götz for the XRD, H. Bender for technical support, C. Busch for the EDX measurements, and A. Ivanov from the Institute Laue Langevin for technical assistance during the neutron diffraction measurements. We also acknowledge support from the DFG within the Schwerpunktprogramm 1458, Grant No. BO3537/1-1.

**References**

1. J. G. Guo, S. F.Jin, G. Wang, S. C.Wang, K. X. Zhu, T. T. Zhou, M. He, and X. L. Chen, Phys. Rev. B **82**, 180520(R) (2010)
2. Z. Shermadini, A. Krzton-Maziopa, M. Bendele, R. Khasanov, H. Luetkens, K. Conder, E. Pomjakushina, S. Weyeneth, V. Pomjakushin, O. Bossen, and A. Amato, Phys. Rev. Lett. **106**, 117602 (2011).
3. R. H. Liu, X. G. Luo, M. Zhang, A. F. Wang, J. J. Ying, X. F. Wang, Y. J. Yan, Z. J. Xiang, P. Cheng, G. J. Ye, Z. Y. Li, and X. H. Chen, Europhys. Lett. **94**, 27008 (2011).
4. D. M. Wang, J. B. He, T.-L. Xia, and G. F. Chen, Phys. Rev. B **83**, 132502 (2011).
5. Z. Wang, Y. J. Song, H. L. Shi, Z. W. Wang, Z. Chen, H. F. Tian, G. F. Chen, J. G. Guo, H. X. Yang, and J. Q. Li, Phys. Rev. B **83**, 140505(R) (2011).
6. A. Ricci, N. Poccia, B. Joseph, G. Arrighetti, L. Barba, J. Plaisier, G. Campi, Y. Mizuguchi, H. Takeya, Y. Takano, N. Lal Saini and A. Bianconi, Supercond. Sci. Technol. **24**, 082002 (2011).
7. J. J. Ying, X. F. Wang, X. G. Luo, A. F. Wang, M. Zhang, Y. J. Yan, Z. J. Xiang, R. H. Liu, P. Cheng, G. J. Ye, and X. H. Chen, Phys. Rev. B **83**, 212502 (2011).
8. A. F. Wang, J. J. Ying, Y. J. Yan, R. H. Liu, X. G. Luo, Z. Y. Li, X. F. Wang, M. Zhang, G. J. Ye, P. Cheng, Z. J. Xiang, and X. H. Chen, Phys. Rev. B **83**, 060512(R) (2011).
9. A. Krzton-Maziopa, Z. Shermadini, E. Pomjakushina, V. Pomjakushin, M. Bendele, A. Amato, R. Khasanov, H. Luetkens, and K. Conder, J. Phys.: Condens. Matter **23**, 052203 (2011).

10. V. Tsurkan, J. Deisenhofer, A. Günther, H.-A. Krug von Nidda, S. Widmann, and A. Loidl, Phys. Rev. B **84**, 144520 (2011).

11. X. G. Luo, X. F. Wang, J. J. Ying, Y. J. Yan, Z. Y. Li, M. Zhang, A. F. Wang, P. Cheng, Z. J. Xiang, G. J. Ye, R. H. Liu, and X. H. Chen, New J. Phys. **13**, 053011 (2011).

12. Z. Shermadini, H. Luetkens, R. Khasanov, A. Krzton-Maziopa, K. Conder, E. Pomjakushina, H-H. Klauss, and A. Amato, arXiv:1111.5142v1.

13. Y. Mizuguchi, H. Takeya, Y. Kawasaki, T. Ozaki, S. Tsuda, T. Yamaguchi, and Y. Takano, Appl. Phys. Lett. **98**, 042511 (2011).

14. J. T. Park, G. Friemel, Yuan Li, J.-H. Kim, V. Tsurkan, J. Deisenhofer, H.-A. Krug von Nidda, A. Loidl, A. Ivanov, B. Keimer, and D. S. Inosov, Phys. Rev. Lett. 107, 177005 (2011).

15. .Y. J. Yan, M. Zhang, A. F. Wang, J. J. Ying, Z. Y. Li, W. Qin, X. G. Luo, J. Q. Li, Jiangping Hu and X. H. Chen, arXiv:1104.4941.

16. F. Ye, S. Chi, Wei Bao, X. F. Wang, J. J. Ying, X. H. Chen, H. D. Wang, C. H. Dong, and Minghu Fang, Phys. Rev. Lett. 107, 137003 (2011).

17. Bao Wei, Huang Qing-Zhen, Chen Gen-Fu, M. A. Green, Wang Du-Ming, He Jun-Bao, and Qiu Yi-Ming, Chin. Phys. Lett. 28, 086104 (2011).